\documentclass[prl,twocolumn,showpacs,letterpaper,showpacs,superscriptaddress,nofootinbib]{revtex4-1}
\usepackage{graphicx,amsmath,amssymb,amsfonts,latexsym,color,dcolumn,bm,subfigure,ulem}

\newcommand{\rd}[1]{\textcolor{red}{#1}}
\newcommand{\bl}[1]{\textcolor{blue}{#1}}

\begin{document}

\title{Stronger limits on hypothetical Yukawa interactions in the  30--8000 nm range}
\author{Y.-J. Chen}
\affiliation{Department of Physics, Indiana University-Purdue University Indianapolis, Indianapolis, Indiana 46202, USA}
\affiliation{Section 3, N16 Process Integration Department 1, Taiwan Semiconductor
Manufacturing Company, HsinChu 30078, Taiwan}
\author{W. K. Tham}
\affiliation{Department of Physics, Indiana University-Purdue University Indianapolis, Indianapolis, Indiana 46202, USA}
\author{D. E. Krause}
\affiliation{Physics Department, Wabash College, Crawfordsville, Indiana 47933, USA}
\affiliation{Department of Physics and Astronomy, Purdue University, West Lafayette, Indiana 47907, USA}
\author{D. L\'opez}
\affiliation{Center for Nanoscale Materials, Argonne National Laboratories, Argonne, Illinois 60439, USA}
\author{E. Fischbach}
\affiliation{Department of Physics  and Astronomy, Purdue University, West Lafayette, Indiana 47907, USA}
\author{R. S. Decca}
\email{Electronic address: rdecca@iupui.edu}
\affiliation{Department of Physics, Indiana University-Purdue University Indianapolis, Indianapolis, Indiana 46202, USA}

\date{ \today}

\begin{abstract}
We report the results of new differential force measurements between a test mass and rotating source masses of gold and silicon to search for forces beyond Newtonian gravity at short separations.  The technique employed subtracts the otherwise dominant Casimir force at the outset and,  when combined with  a lock-in amplification technique,  leads to a significant improvement (up to a factor $10^{3}$) over existing limits on the strength (relative to gravity) of a putative force in the 40--8000 nm interaction range.
\end{abstract}


\maketitle

Although the gravitational attraction between two point masses was the first force to be described it remains, in comparison to other fundamental forces, poorly characterized. Unification theories, such as string theory, which introduce  $n$ compact extra spatial dimensions, predict deviations from Newtonian gravity over sub-mm scales \cite{ADD98,Adelberger03}. Also, many extensions to the Standard Model predict the existence of new light bosons, the exchange of which would lead to new forces. In both cases, the existence of compact extra dimensions or exchange of new light bosons, the non-Newtonian interaction between two point masses $m_{1}$ and $m_{2}$ separated by a distance $r$  can be parametrized as 
\begin{equation}
V(r) = -G \frac{m_{1}m_{2}}{r}\alpha e^{-r/\lambda},
\label{Vnewt}
\end{equation}
where $G$ is the Newtonian gravitational constant, $\alpha$ is the strength of the Yukawa-like correction arising from new physics, and $\lambda$ is its characteristic range. In the case of compact extra dimensions, $\lambda$ closely corresponds to the size of the extra-dimension. For the  exchange of a boson of mass $m$, $\lambda= \hbar/mc$ \cite{Ephraim book}. 

Motivated in part by these considerations a  large number of experiments have been conducted  to constrain the value of $\alpha$ (see for example the reviews \cite{Review09, Review14}). While they  have been successful in constraining $|\alpha| < 1$ for $\lambda > 50~\mu$m \cite{Kapner07}, the limits on $\alpha$ are much less restrictive for  $\lambda < 10~\mu$m. Constraints on $\alpha$ for small values of $\lambda$ are much more difficult to achieve due to the small effective masses (i.e. mass within a distance $r \sim \lambda$ of the surface) interacting through the Yukawa-like contribution. Compounding the problem at  sub-micron separations, the effects of vacuum fluctuations eventually become  dominant after electrostatic contributions have been minimized. Hence, many of the limits in the $\lambda \in [10,10000]~$nm  range have been obtained by subtracting from the measured interaction the calculated contribution from  the Casimir force \cite{Ederth00, Harris00,Decca07,Sushkov11} which arises from vacuum fluctuations. While useful, this approach has  two main drawbacks: (i) The subtracted background is relatively  large, and hence small corrections to the background result in large changes in the derived limits; (ii) It is not clear what the appropriate  background to subtract is.  While some groups use a plasma model for the extrapolation to zero frequency of the dielectric function of the metal, others use a Drude model \cite{KMM}. The correct approach remains a matter of controversy, and new experiments have been proposed to help resolve this problem \cite{Bimonte}.

In the absence of electromagnetic contributions, a comparison of the forces exerted on a test mass by materials of different densities leads to constraints on $\alpha$ and $\lambda$ in Eq.~(\ref{Vnewt}).  Different materials differ not only in their densities but also in their response to vacuum fluctuations, and hence these effects must be suppressed when searching for the presence of putative new forces at sub-micron separations.  The ``isoelectronic'' or ``Casimir-less'' technique introduced in \cite{Decca05} capitalizes on the fact that the response of a sample to vacuum fluctuations is mainly a surface effect, whereas any new  force interacts  with a portion within range $\sim \lambda$ of its surface.  In the ``Casimir-less'' technique contributions from vacuum fluctuations are suppressed by coating the source mass with a layer of Au of thickness larger than  the plasma wavelength of Au, $\lambda_{\rm p} = 135$~nm such that the difference in the Casimir interaction between the underlying structure in the source mass with the test mass is attenuated by a factor larger than $10^6$ \cite{Decca05,Matloob01}.
The Au layer thus serves not only to reduce conventional electrostatic effects as in other experiments but, more significantly, to suppress vacuum fluctuation contributions associated with the composition of the test mass.

While earlier experiments successfully demonstrated the possibility of subtracting the Casimir background, the performance of the technique was limited by two experimental constrains.
(i) To observe a signal at the resonance frequency $\omega_r$ of the mechanical oscillator (without anything moving at $\omega_r$) a  heterodyne technique was used. The test mass was harmonically positioned over the two sides of the source mass at $\omega_1$  while the separation between the test and source masses was harmonically varied with amplitude $\delta z$ at $\omega_2 = \omega_r- \omega_1$, effectively reducing  the hypothetical Yukawa-like signal  by $\delta z/\lambda \sim 0.02$.  
(ii) The sample was made in such a way that  the thicknesses of the two sides of the  source mass were unintentionally different. This translated into a  $\sim$ 3~fN systematic signal identified with the distance dependence of the Casimir force. This residual signal yielded the limits obtained in Ref.~\cite{Decca05}. 

In this paper we report a new approach to improve the limits in the $\{\lambda, \alpha\}$ phase space. The use of a rotating source mass allowed us to fully utilize the high force sensitivity provided by the large mechanical quality of the microelectromechanical torsional oscillator (MTO)\cite{Lopez05}. Furthermore, an implementation of the source mass where there is no correlation between its thickness and its angular position yielded an unprecedented level of subtraction of the  background arising from vacuum fluctuations. 

The test mass (a $R=149.3 \pm 0.2~ \mu$m sapphire sphere covered with a $t_{\rm Cr} \sim 10$~nm layer of Cr and a $t_{\rm Au} \sim 250$~nm Au-film) was glued close to the edge (at a distance $b = 235\pm 4~ \mu$m from the axis of rotation) of the $500~ \mu$m $\times~500~ \mu$m  plate of the oscillator. Gluing the sphere reduced the MTO's natural frequency of oscillation from $f_{0} = 708.23 \pm 0.05$~ Hz to $f_r = 307.34 \pm 0.05$~ Hz, and it reduced the oscillator's quality factor from $Q \sim 9000$ to $Q\simeq 7200$ for a pressure $P \leq 10^{-5}$~torr. The experiments were performed at $P \simeq 10^{-5}$~torr and the motion of the plate was detected by the change in capacitance between the plate and the underlying electrodes as in \cite{Decca05,Decca09,Decca11}. Calibration of the MTO was performed by using the electrostatic interaction between the Au-coated test and source masses \cite{Decca09}. The calibration was performed with the source mass stationary, and the distance was monitored and measured using a two-color interferometer (with a sensitivity of 0.2~nm). After performing the calibration, the potential difference between the sphere and plate was adjusted to minimize the electrostatic interaction. With this MTO a thermally limited minimum detectable force $F_{\rm min}(f_r) \sim 6~{\rm fN}/\sqrt{\rm Hz}$ was calculated when working at resonance at 300 K\cite{Resonance}. Since  $f_r$ is a function of separation due to the non-linear nature of the Casimir interaction,  it was continuously monitored.  

\begin{figure}[b]
\vspace{-0.650cm}
\centerline{\includegraphics[width=7.0cm]{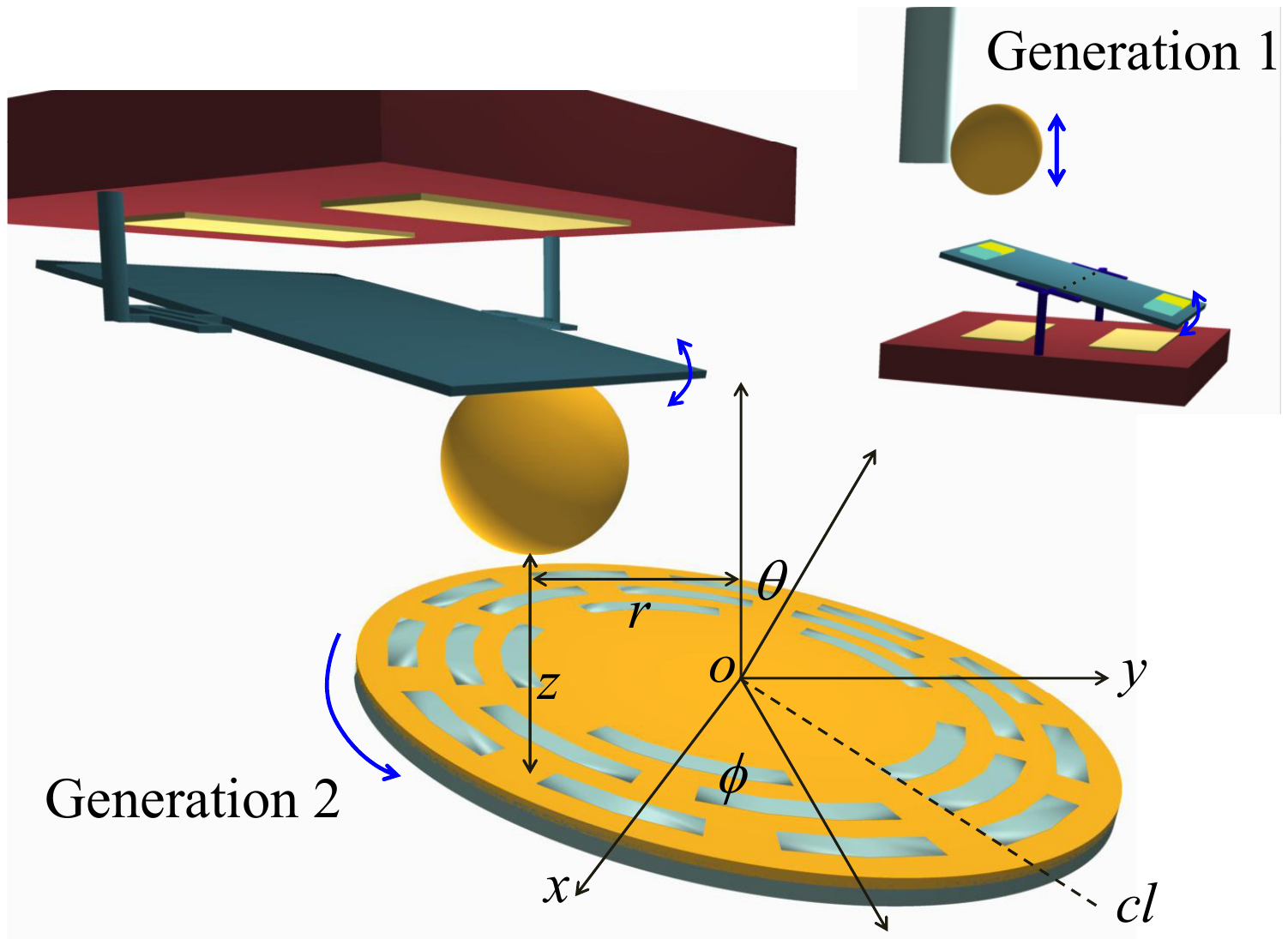}}
\vspace{-0.8cm}
\caption{(Color online) Schematic of the experimental setup (not to scale). The Au-coated sphere is glued to the oscillator. Three regions with $n = 5, 8, 11$ Au-Si sectors are shown. The actual sample has $n = 50, 75, \cdots, 300$. The $\{x,y\}$ plane defines the plane of rotation of the spindle. $cl$ is the line where all the different regions with Au-Si sectors coincide. $\theta$ is the  instantaneous axis of rotation, $\phi =\omega t$ is the angle of rotation. The distance $z$ is measured from the vertex of the spherical test mass to the source mass. $r$ is the distance from the vertex of the test mass to the center of the source mass, $o$. Displacements $\Delta r$ between $o$ and the axis of rotation are not shown for clarity. For comparison, a schematic of the setup used in \cite{Decca05} is shown.}
\label{fig1}
\end{figure}

 A five axis stepper-motor-driven positioner and a three axis piezoelectrically driven system were used to bring the test mass in close proximity ($z \in [200,1000]$~nm) to the source mass. The source mass was fabricated by depositing a $d_{\rm Cr} = 10$~nm thick layer of Cr on a 1~inch diameter $100~\mu$m thick [100] oriented Si wafer. A $d_{\rm tm} = 2.10 \pm 0.02 ~\mu$m thick layer of Si was deposited on top of the Cr covered Si wafer. Using conventional photolithography, a photoresist structure  consisting of concentric sectors  was defined in the Si. The Si not covered by the photoresist was removed down to the Cr layer using CF$_4$ reactive ion etching. After removing the photoresist, Au was thermally evaporated and the structure mechanically polished to expose the Si sectors. This process defined a structure with a surface consisting of a center circle of Au with a radius $R_1 = 4$~mm, then a $200~\mu$m wide ring with  50 sectors of Au/Si, and a $150~\mu$m wide Au ring. The sequence of $200~\mu$m wide rings with Au/Si sectors and  $150~\mu$m wide Au rings was repeated with the number of Au/Si sectors increasing by 25 for each concentric ring until the last one with 300 sectors, which was located at $R_{11}=7.5$~mm. This structure was glued with NOA61 UV curing cement to a BK7 Schott glass flat with the original Si wafer exposed. The wafer was etched away using KOH, and then a $d_{\rm Au} =150 \pm 3~$nm layer of Au was deposited by thermal evaporation. The exposed Au surface was characterized by white light interferometry (WLI) and atomic force microscopy (AFM), which showed an optical quality film with no memory  of the underlying structure. The  1024 $\times$ 1024 AFM images obtained over different  $10~ \mu$m $\times 10~ \mu$m regions yielded position-independent 60~nm peak-to-peak topographic roughness. Excluding a few  isolated spikes $\sim 50$~nm tall and about 100~nm across, the sample has a rms roughness of 1.5 nm. The disk was then mounted on an air bearing spindle. It was optically verified that the center of the disk and the axis of rotation of the spindle coincided to better than $\Delta r \sim ~10~\mu$m. The flatness and alignment of the sample were checked {\it in-situ} using a fiber interferometer (response time 10 ms). It was found that the surface of the sample was perpendicular to the axis of rotation to better than $z_{0}$~=~20~nm at $R_{11}$ when rotating the disk at $\omega = 2\pi$~rad/s. 

The air bearing spindle worked under a constant air flow of several liters/min. The top of the source mass was at a distance $D = 4$~cm from the air exhaust. To prevent air leaks into the chamber, the spindle was mounted with a circular skirt which rotated with the spindle. The seal between the skirt and the vacuum chamber was provided by high molecular weight oil. Oil contamination inside the chamber was precluded using chilled water refrigeration ($T= 10^{\circ}$ C) on a system of baffles and traps.

With the sphere placed at $R_i +100~\mu$m (with $n$ Au/Si sectors) the air bearing spindle was rotated at $\omega_r=2\pi f_r/n$. In this manner, a force arising from the potential given in Eq.~(\ref{Vnewt}) would have manifested itself at $f_r$ even though  there were no parts moving at $f_r$. The Newtonian gravitational attraction between the sphere and the structured sample yields a force  $F_N \sim 10^{-20}$~N, undetectable by our system. Hence an integration of only the Yukawa-like part of Eq.~(\ref{Vnewt}) over the geometry of the sample is necessary. Disregarding finite size effects across the width of the ring \cite{Decca11}, the expected difference when the sphere is over a Au or Si sector is \cite{Decca05,Decca PFA}

\begin{eqnarray}
\Delta F_{h}(z)& = &-4\pi^2 G\alpha\lambda^3e^{-z/\lambda}R K_t  K_s, \label{force}\\
K_t& =
&\left[\rho_{\rm Au}-\left(\rho_{\rm Au}-\rho_{\rm Cr}\right)e^{-t_{\rm Au}/\lambda} \right. \nonumber \\ & &\left. -\left(\rho_{\rm Cr}-\rho_{\rm s}\right)e^{-(t_{\rm Au}+t_{\rm Cr})/\lambda} \right], \nonumber \\
K_s&=&\left[\left(\rho_{\rm Au}-\rho_{\rm Si}\right) e^{-\left(d_{\rm Au}+d_{\rm Cr}\right)/\lambda}
\left(1-e^{-d_{\rm tm}/\lambda}\right)\right], \nonumber
\label{Fhyp}
\end{eqnarray}

\noindent where $K_t$ ($K_s$) is a term associated only
with the layered structure of the test (source) mass, $\rho_{\rm s}$, $\rho_{\rm Cr}$,  $\rho_{\rm Au}$,  and $\rho_{\rm Si}$
 are the sapphire,  Cr, Au and Si densities respectively.
   
\begin{figure}[htb]
\vspace{-.3cm}
\centerline{\includegraphics[width=9cm]{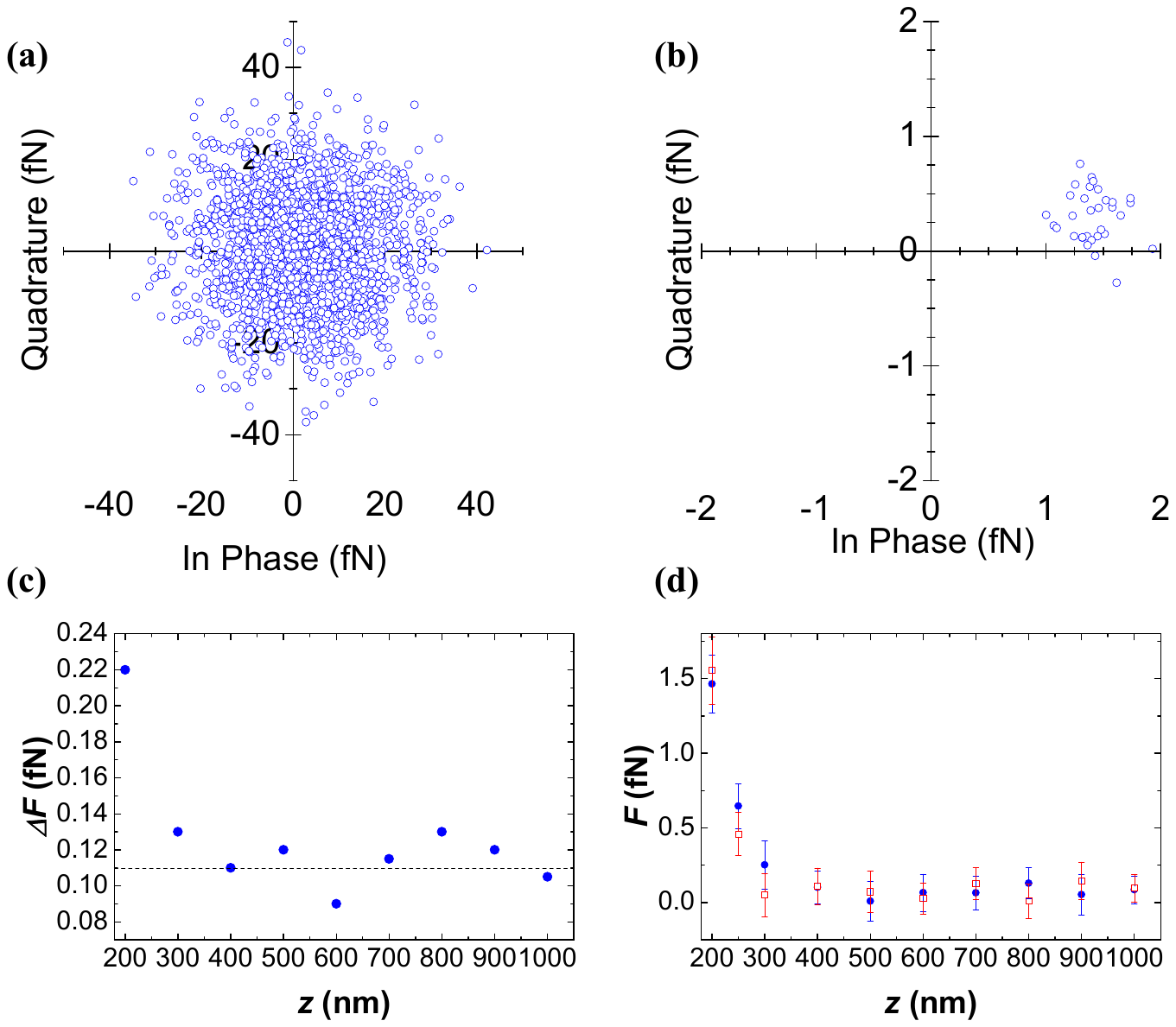}}
\vspace{-0.5cm}
\caption{(Color online) (a) Lock-in amplified detection of the signal with an integration time $\tau = 1$~s at a separation $z = 200$~nm. (b) Same as in (a) for $\tau =3000$~s. (c) Standard deviation for 15 different realizations of the experiment with $\tau = 3000$~s as a function of separation. The horizontal dashed line is the statistical noise in the amplitude of the oscillator at $T = 293$~K (which was also measured over $\tau = 12,000$~s at $z= 3~\mu$m). (d) Measured interaction as a function of separation. The two data sets were obtained on top of the region with $n=300$ (\bl{ $\bullet$}) and over a section of the sample without Si at a radius $r = 8$~mm (\rd{ $\Box$}). The error bars represent the standard deviation for 10 repetitions with $\tau=3000$~s. }
\label{fig2}
\end{figure}

The  setup is optimized to select the first harmonic of the force associated with the angular distribution of the sample.  Other harmonics and all forces with different angular dependences are outside of the resonance peak of the MTO and consequently ``filtered'' by the sharp $\Delta f \simeq 40$~mHz resonance peak of the oscillator.

Results obtained by doing the experiment over the $n=300$ ring are shown in Fig.~\ref{fig2}. These results were obtained by using a lock-in detection technique at $f_r$. A mark on the outside of the source mass coincident with the {\it cl} line in Fig.~\ref{fig1})  was used to define the origin of the phase. Many features are worth  noting: (i) Increasing the integration time $\tau$ decreases the random noise of the measured force, as expected. (ii) At separations $z \leq 300$~nm the statistical noise is larger than the minimum detectable force. This happens in a region where the Casimir force is large. At larger separations, the statistical noise is close to the minimum detectable force.  It was also observed that this noise was independent of the angular frequency of the disk (for $\frac{\omega}{2\pi} \in [0.2, 20]$~Hz). (iii) There is a separation-dependent net force measured, which is practically independent of the radial position of the sphere over the rotating disk. This signal increases proportionally to $\omega$ as the disk is rotated at higher subharmonics of $\omega_r =2\pi f_r$. 

Point (iii) indicates an incomplete subtraction of background forces. Kelvin probe force microscopy was performed in the sample over different $5\times5~\mu$m$^2$ regions. This contribution is not of electrostatic origin. It was observed that the main potential islands had a characteristic size $\ell \sim 200$~nm with $V_{\rm rms} < 5$~mV. Since the experimental system provides a sharp filtering of the signal at $\omega_r$, this implies the electrostatic contribution to the signal at $\omega_r$ would be undetectable,  $\Delta F_{\rm el}(z) \leq 10^{-17}$~N \cite{Behunin14}. Furthermore, if detected, the electrostatic signal would have a radial dependence when the disk is rotated at constant frequency \cite{Behunin14}, which was not observed. 

Variations in the separation between test and source masses could yield the observed background through the separation dependence of the Casimir force $F_{\rm C}(z) \,\propto\, z^{-\alpha}$ \cite{KMM}.  $F_{\rm C}(200~{\rm nm})=34~ {\rm pN}$ and $\alpha = 2.78$ were experimentally determined in the actual  configuration. The observed signal at  $\omega_r= n \omega$ must appear through $z(t)$.

The observed signals are consistent with the axis of rotation of the spindle having both an impulse-like $\Delta \theta_1$ once per revolution   and a random wobbling. The impulse-like wobble has been identified by analyzing its frequency dependence. Varying the frequency of the  spindle the harmonic components of the signal observed at $\omega_r$ are  consistent with a once per revolution impulse-like signal. The random wobbling is observed to have white-noise characteristics, $\langle \theta_2(\tau) \theta_2(0) \rangle = \Theta^2 \delta(\tau)$, where $\Theta$ is a constant, in the range of frequencies investigated, between 0.1 and 20 Hz. This random noise increases the minimum detectable force from $\sim$ 6 to $\sim$ 12 fN/$\sqrt{\rm Hz}$ at $z =200$~nm. Associated with any $\Delta \theta$, there is a change in separation $\delta z \sim D \delta \theta$ between the sphere and the rotating sample, which induces a change in the Casimir force. Since in our experiment $D \sim 4$~cm, it follows that $\Theta \sim 5\times 10^{-10}$~rad and $\Delta \theta_1 \leq 10^{-7}$~rad. Neither of these angular deviations can currently be measured directly.

\begin{figure}[thb]
\vspace{-0.5cm}
\centerline{\includegraphics[width=9cm]{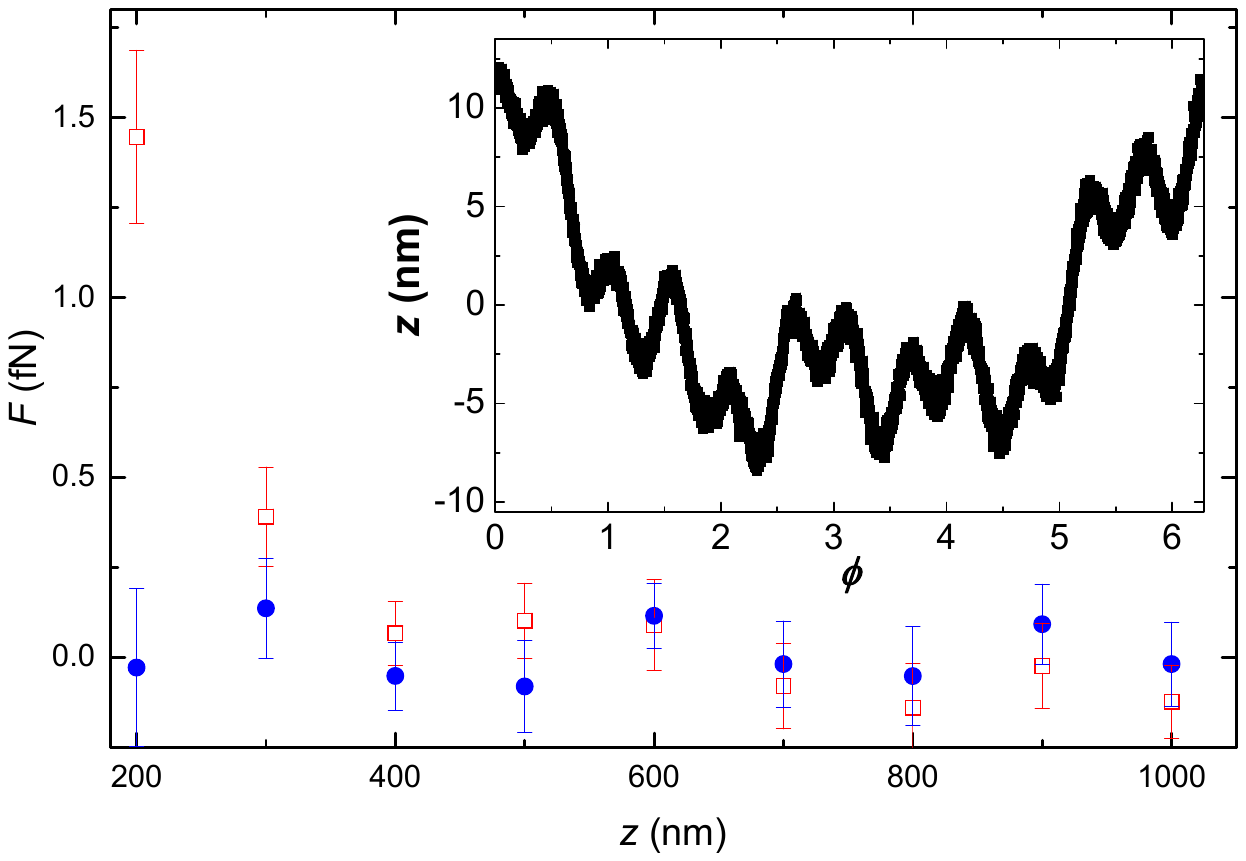}}
\vspace{-2cm}
\caption{(Color online) Measured interaction as a function of separation obtained on top of the region with $n=300$. (\bl{ $\bullet$}) quadrature signal; (\rd{ $\Box$}) in phase signal (see text). Errors represent the standard deviation for 10 repetitions with $\tau = 3000$~s. Inset: The sample's topography, $z(\phi)$ over the $R_{11}$ circle, obtained by white light interferometry. }
\label{fig3}
\end{figure}

The lack of parallelism between the normal to the disk and the rotational axis leads to the time-varying separation $z(t) = z_{s}+z_{0}\cos(\omega t)$.  Its contribution $\Delta F_{\rm par}$ at $\omega_r$ enters through the nonlinear dependence of the Casimir force on the separation. A Taylor's expansion of $F_{\rm C}(z) \,\propto\, z^{-\alpha}$ shows that the contribution at $\omega_r$ is attenuated by $\sim (z/z_{0})^n$, making it unobservable for all $n$ in our setup.
 This is also  the case for precession of the spindle. 
Similarly, the lack of flatness of the sample  generates the angular-position-dependent separation $z(\phi)$, which was measured using white light interferometry (see inset in Fig.~\ref{fig3}). Inserting these data into  $F_{\rm C}(z)$ yields a contribution $\Delta F_{\rm top} < 0.05~{\rm fN}$.

The effect of the once-per-revolution $\Delta\theta_1$  was minimized in the following way: $\Delta\theta_1$ happens  at a characteristic $\phi_o$. As the disk is positioned for the first time on the spindle, there is an unknown angle $\phi_x$ between the line {\it cl} and the line defined by $\phi_o$. When the sphere is positioned over a region with only  Au in the source mass, a signal with a non-zero phase is detected (recall the zero-phase is defined at {\it cl}). The sample is then very carefully repositioned over the disk until the phase of the detected signal is zero. In this situation $\phi_x = 0$ is assumed.

An approach where the zero of the phase is redefined could also be used,  however the method of repositioning the sample is superior to redefining the zero of the phase. While it is expected that the signal described by Eq.~(\ref{force}) is in quadrature (i.e. it should be an odd signal with respect to $\phi$), in principle, it is not known if the model is correct. The data shown in Fig.~\ref{fig3} were obtained in this manner. Furthermore, the same component of the signal in phase was measured with the sphere placed at any radii.  It was  determined that the hypothetical force (in quadrature) is consistent with zero within the experimental error for any radii $R_n$.

The $\Delta r$ shift between the axis of rotation and the center of the source mass would also yield a signal at $\omega$ if $\Delta F_{h}$ were observable, although it would be  attenuated by $R_i/\Delta r$  at $\omega_r$. Similarly,  finite size effects as the Au/Si interface of the source mass moves under the test mass are negligible at $\omega_r$ when compared with the statistical errors.

\begin{figure}[htb]
\vspace{-0.8cm}
\centerline{\includegraphics[width=9cm]{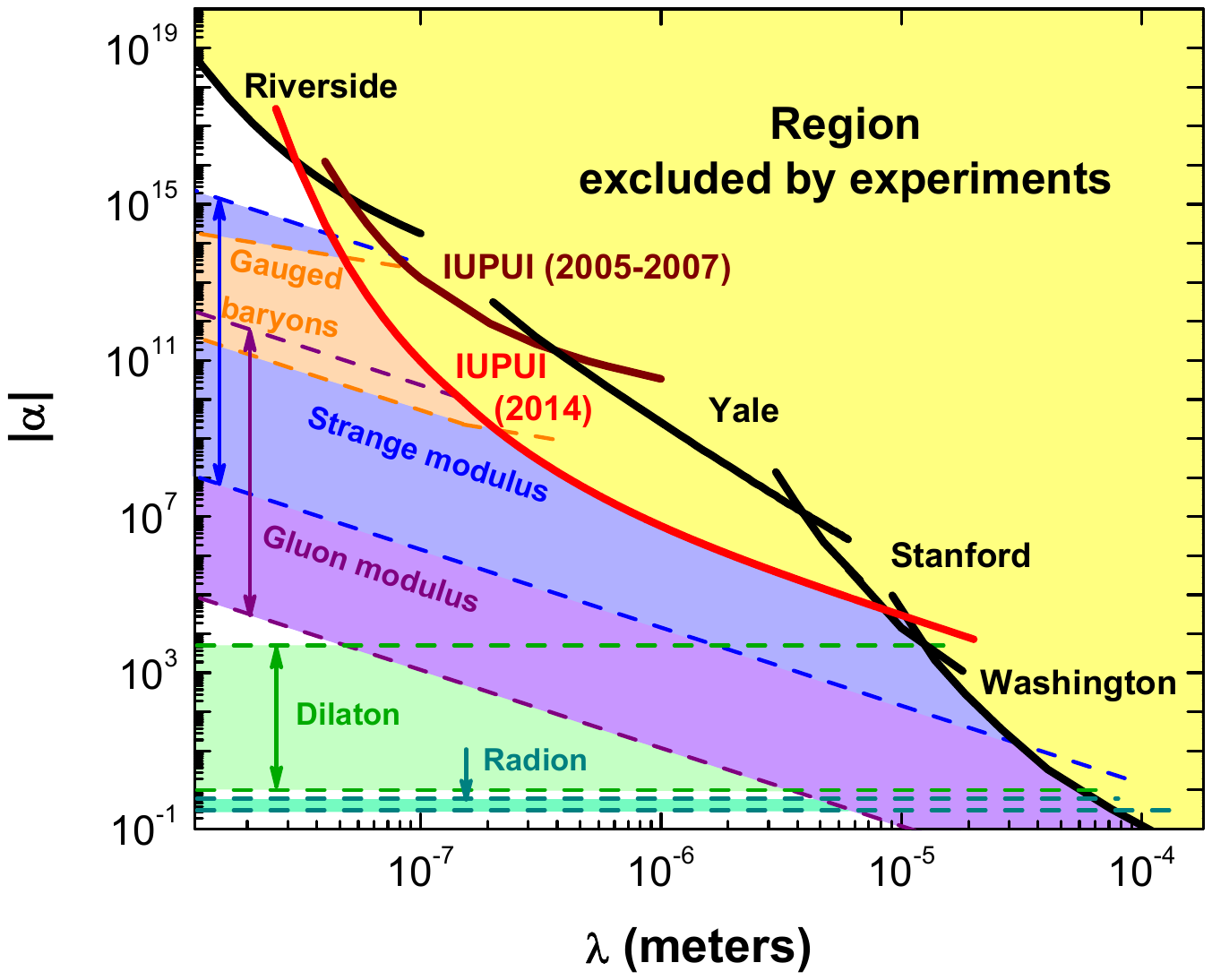}}
\vspace{-.5cm}
\caption{(Color online) Values in the ${\lambda,\alpha}$ phase-space excluded by experiments. The red curve represents the limits obtained in this work. Previous limits from Riverside \cite{MohideenLim}, IUPUI \cite{Decca07,Decca05}, Yale \cite{LamoreauxLim}, Stanford \cite{StanfordLim}, Washington \cite{WashingtonLim} and  theoretical predictions \cite{ADD98,Adelberger03,Kaplan,Antoniadis,Yang} are also shown.}
\label{fig4}
\end{figure}

While the relevant error is  $\Delta F_{\rm rand}$, the overall error was obtained as an addition of the random and systematic errors $\Delta F = \Delta F_{\rm rand} + \Delta F_{\rm syst}$. The individual systematic errors described in this paper were considered to be independent to obtain $\Delta F_{\rm syst}$. $F(z)$ in Fig.~\ref{fig3} associated with the hypothetical force  is consistent with zero and was used to  establish new limits in $\{\lambda,\alpha\}$ space at the 95\% confidence level. The envelope of the curves where the first harmonic of $F_h(z)$ is compared with max$\{|F(z)+2\Delta F(z)|$,$|F(z)-2\Delta F(z)|\}$,  is shown in Fig.~\ref{fig4}.  The new limits obtained with the introduced ``Casimir-less'' measurement technique represent a significant improvement over previous experiments:  new boundaries  have been established in a spatial range covering more than 3 orders of magnitude ($\lambda \in [30,8000]$~nm), with improvements as large as $10^3$ in Yukawa-like corrections to Newtonian gravity at $\lambda = 300$~nm. 

This work was performed, in part, at the Center for Nanoscale Materials, a U.S. Department of Energy, Office of Science, Office of Basic Energy Sciences User Facility under Contract No. DE-AC02-06CH11357.  R.S.D. acknowledges support from the National Science Foundation through Grant
PHY-0701636 and from Los Alamos National Laboratory through Contract 49423-
001-07. R.S.D. is also indebted to the  IUPUI Nanoscale Imaging Center, the IUPUI Integrated Nanosystems Development Institute, and the Indiana University Center for Space Symmetries for financial and technical support. The work of  E.F. was supported in part by U.S. Department of Energy contract DE-AC02-76ER071428.


\begin{references}
\bibitem{Resonance} Away from resonance, the electronic background imposes a constraint on the minimum detectable force. In our case it is $\sim$~0.5~pN.
\bibitem{adjust} It was verified that the same results were obtained by a simple change of phase in the lock-in detection scheme.
\end{references}
\end{document}